\newif\ifanonymousversion
\anonymousversionfalse

\ifanonymousversion

\documentclass[conference, screen, anonymous]{IEEEtran}

\else

\documentclass[conference, screen]{IEEEtran}

\fi

\usepackage{tikz}
\usepackage{amsmath}
\usepackage{xcolor}
\usepackage{filecontents}

\usepackage{hyperref}
\usepackage{soul}
\usepackage[braket, qm]{qcircuit}
\usepackage{physics}
\usepackage{enumitem}
\usepackage{multirow}
\usepackage{booktabs}
\usepackage{algorithmicx}
\usepackage{algpseudocode}
\usepackage{algorithm}

\usepackage{caption}
\usepackage{subcaption}

\usepackage{balance}
\usepackage{nicematrix}

\usepackage{subcaption}

\usepackage{amsfonts}
\usepackage{flushend}
\usepackage{url}

\newcommand\scalemath[2]{\scalebox{#1}{\mbox{\ensuremath{\displaystyle #2}}}}

\newcommand{\nsf}[1]{\href{https://www.nsf.gov/awardsearch/showAward?AWD_ID=#1}{#1}}

\IEEEoverridecommandlockouts

\begin{document}

\date{}

\title{A Thorough Study of State Leakage Mitigation in Quantum Computing with One-Time Pad\thanks{This work was supported in part through NSF grant \nsf{2312754}.}} 

\ifanonymousversion

\author{Anonymous Submission, HOST 2023}

\else

\author{
\IEEEauthorblockN{Chuanqi Xu}
\IEEEauthorblockA{\textit{Dept. of Electrical Engineering} \\
\textit{Yale University}\\
New Haven, CT, USA \\
chuanqi.xu@yale.edu}
\and
\IEEEauthorblockN{Jamie Sikora}
\IEEEauthorblockA{\textit{Dept. of Computer Science} \\
\textit{Virginia Polytechnic Institute and State University}\\
Blacksburg, VA, USA \\
sikora@vt.edu}
\and
\IEEEauthorblockN{Jakub Szefer}
\IEEEauthorblockA{\textit{Dept. of Electrical Engineering} \\
\textit{Yale University}\\
New Haven, CT, USA \\
jakub.szefer@yale.edu}
}

\fi

\maketitle

\begin{abstract}
The ability for users to access quantum computers through the cloud has increased rapidly in recent years. Despite still being Noisy Intermediate-Scale Quantum (NISQ) machines, modern quantum computers are now being actively employed for research and by numerous startups. Quantum algorithms typically produce probabilistic results, necessitating repeated execution to produce the desired outcomes. In order for the execution to begin from the specified ground state each time and for the results of the prior execution not to interfere with the results of the subsequent execution, the reset mechanism must be performed between each iteration to effectively reset the qubits. However, due to noise and errors in quantum computers and specifically these reset mechanisms, a noisy reset operation may lead to systematic errors in the overall computation, as well as potential security and privacy vulnerabilities of information leakage. To counter this issue, we thoroughly examine the state leakage problem in quantum computing, and then propose a solution by employing the classical and quantum one-time pads before the reset mechanism to prevent the state leakage, which works by randomly applying simple gates for each execution of the circuit. In addition, this work explores conditions under which the classical one-time pad, which uses fewer resources, is sufficient to protect state leakage. Finally, we study the role of various errors in state leakage, by evaluating the degrees of leakage under different error levels of gate, measurement, and sampling errors. Our findings offer new perspectives on the design of reset mechanisms and secure quantum computing~systems.
\end{abstract}

\section{Introduction} 

The term Noisy Intermediate-Scale Quantum (NISQ) quantum computer is used to refer to the current quantum computers~\cite{preskill2018quantum}. 
Despite already having promising applications in optimization, chemistry, and other crucial fields~\cite{lanyon2010towards,jones1998implementation,mermin2007quantum}, today's NISQ quantum computers are still too limited to provide quantum error correction~\cite{Devitt_2013}, and execute ``large'' algorithms, such as Shor's algorithm~\cite{365700} and Grover's algorithm~\cite{10.1145/237814.237866}. 
However, NISQ quantum computers are being developed quickly; 433-qubit machines are now available, and above 4000-qubit is anticipated soon~\cite{433qubit}. 

Nowadays, quantum computers from various suppliers are already accessible through cloud-based services such as IBM Quantum~\cite{ibm_quantum}, Amazon Bracket~\cite{braket}, and Microsoft Azure~\cite{azure}. 
Without having to buy or maintain them, remote access makes it simpler to run algorithms on actual quantum computers, but also leads to privacy and security concerns with such open access. 
For example, malicious users can try to gather the leaked information to learn the state of the victim user's qubits through the victim's results.
One possible source of state leakage is the noisy operation, such as the reset operation which is necessary between circuit executions to reset qubits. 
As a result of noisy and erroneous reset operations, information may be carried over to subsequent executions, and this leakage may be abused by attackers. 
Such a weakness is shown in reset attacks~\cite{10.1145/3548606.3559380}, side-channel attacks~\cite{9951250}, and higher-energy state attacks~\cite{10.1145/3576915.3623104}. 
Given that the information is leaked sequentially from earlier executions to later executions, this type of state leakage can be referred to as ``horizontal" leakage. 
On the other hand, ``vertical'' leakage, which simultaneously occurs from qubits to qubits, is another kind, which is demonstrated in crosstalk attack~\cite{10.1145/3370748.3406570, 9193969, 9840181, 10133711} and qubit sensing~\cite{saki2021qubit}.

The one-time pad (OTP) is a well-known, powerful tool in cryptography to perfectly encrypt information~\cite{doi:10.1080/01611194.2011.583711}. 
The idea is to generate a random key to XOR, or \emph{pad}, the plaintext into a perfectly secure ciphertext. 
While this is proved to be secure in classical computing, it is not enough for quantum computing because of the neglect of some crucial degrees of freedom, such as phase information. 
To extend the classical OTP (COTP) to the quantum setting, the quantum one-time pad (QOTP) has been proposed~\cite{PhysRevA.67.042317, 892142} which can perfectly secure qubits, and thus can be a potential approach to mitigate the information leakage in quantum computing.

In this paper, we aim to thoroughly study \emph{horizontal leakage} by deducing the theoretical model for describing the process of state preparation in the previous execution, reset mechanism between executions, and information collection in the following execution. 
As demonstrated later, state leakage is mainly due to the noise and errors in quantum computers, specifically reset operations. Thus, depending on the implementations and error rates of reset operations, there may be different amounts of state leakage, and commonly used reset operations are analyzed and evaluated in this work. 
As a countermeasure, we propose to apply the OTP to mitigate the state leakage. 
Also, we show that the COTP is sufficient to mitigate leakage in measurement-based reset, nullifying the need for the more expensive QOTP. 
We evaluate our technique on both quantum computers and simulators to secure quantum reset operations.

\section{Background}
\label{background}

This section introduces key concepts in quantum computing.

\subsection{Quantum Computing Basics} 

Analogous to the bit in classical computing, the quantum bit (qubit) is the basic unit in quantum computing. 
A qubit can be represented by a two-dimensional unit complex vector: 
$\ket{\psi} = (\alpha, \beta)^T$, where $|\alpha|^2 + |\beta|^2 = 1$ due to the requirement for unity. 
Any qubit can be expressed as a linear combination 
$\ket{\psi} = \alpha \ket{0} + \beta \ket{1}$. 
where $\ket 0 := (1, 0)^T$ and $\ket 1 := (0, 1)^T$ which can be thought of as the $0$ and $1$ inside of a traditional computer. 
More generally, the state space of $n$-qubit states are spanned by $2^n$ basis states starting from $\ket{0\dots0}$ to $\ket{1\dots1}$, and an $n$-qubit state $\ket{\psi}$ can be represented as 
\begin{equation}
\scalemath{1.0}{
    \ket{\psi} = \sum_{i = 0}^{2^n - 1} a_i \ket i
}
\label{eq:qubit_state}
\end{equation} 

Qubits are controlled and evolve under quantum gates, which can be represented as unitary matrices, i.e. for a quantum gate represented by a matrix $U$, it requires that $UU^\dagger = U^\dagger U = I$, where $U^\dagger$ denotes conjugate transpose. 
Several quantum gates that are used in this paper, and can be executed on today's real quantum computers, are listed below:
\begin{equation}
\scalemath{0.8}{
I = \begin{pmatrix} 
1 & 0 \\ 
0 & 1
\end{pmatrix}, \; 
X = \begin{pmatrix} 
0 & 1 \\ 
1 & 0
\end{pmatrix}, \; 
Z = \begin{pmatrix} 
1 & 0 \\ 
0 & -1
\end{pmatrix}, \; 
XZ = \begin{pmatrix} 
0 & -1 \\ 
1 & 0
\end{pmatrix}
}
\label{eq:gate_matrix}
\end{equation} 
as well as the rotation-by-$\theta$ gate: 
\begin{equation}
\scalemath{0.9}{
R_{\theta} = \begin{pmatrix} 
\cos\frac{\theta}{2} & -i\sin\frac{\theta}{2} \\ 
-i\sin\frac{\theta}{2} & \cos\frac{\theta}{2}
\end{pmatrix}}
\end{equation} 
The above examples are single-qubit gates.  
In general, an $n$-qubit gate can be expressed by a $2^n\times 2^n$ unitary matrix. 
Some multi-qubit gates can create entanglement, which is a phenomenon that cannot be found in the classical world. 
Moreover, a collection of gates is called a \emph{circuit}, which is the form of quantum computation considered in this work. 

At the end of a quantum circuit, the final state can be measured to get computation results. 
According to Born's rule, for a state described as in Equation~\ref{eq:qubit_state}, the probability of measuring or observing $\ket i$ is given by $P(\ket i) = |a_i|^2$. 
Moreover, the measurement leads to the collapse of the quantum state, i.e. if the measurement result is $\ket i$, then the state will collapse to $\ket i$ afterward, a stark contrast to the way in classical computing. 

In addition to the way we introduced quantum states above, which we refer to as \emph{pure states}, we can also have \emph{mixed states}, which is a probability distribution over quantum states. 
Suppose with probability $p_i$ one is given the quantum state $\ket{\psi_i}$, such a mixture is denoted $\{ (p_i, \ket{\psi_i}) \}$ and is represented mathematically using the density matrix 
\begin{equation} 
\rho = \sum_{i} p_i \ketbra{\psi_i}
\end{equation} 
noting that $\bra{\psi_i} := \ket{\psi_i}^\dagger$ and thus $\rho$ is a matrix. 
The probability of measuring $\ket{i}$ is given by $P(\ket{i}) = \bra{i} \rho \ket{i}$. 
If $\rho = \frac{1}{n} I_n$, where $I_n$ is the $n$-dimensional identity matrix, then it is the maximally mixed state, i.e. the probability of measuring any state $\ket{i}$ will be $\frac{1}{n}$.

\subsection{Classical One-Time Pad (COTP)}

Suppose Alice and Bob share a uniformly random bit-string $k \in \{0, 1\}^n$ which is only known to them. 
If Alice has a message $m \in \{0, 1\}^n$ and sends it to Bob $c = m \oplus k$, with $c\in \{0, 1\}^n$ the ciphertext and $\oplus$ the bit-wise XOR, then Bob can recover the message by noting that $m = c \oplus k$. 
However, anyone else who does not know $k$ will see a uniformly random bit-string and thus will have no information about $m$.

Technically, the same thing can be done with a qubit. If Alice has a qubit $\ket \psi$ and share one bit $k$ with Bob, Alice can send $\ket{\psi^\prime} = X^k \ket \psi$ to Bob, where $X$ is the Pauli-$X$ gate in Equation~\ref{eq:gate_matrix}. Then Bob can decrypt the received state to obtain Alice's state by $\ket \psi = X^k \ket{\psi^\prime}$. In the following, we refer to this scheme as the classical one-time pad, or COTP for short.

It functions the same as in classical computing when $\ket \psi = \ket 0 \text{or} \ket 1$, and thus is proved to be secure. However, this scheme is not secure on other occasions. More specifically, for a qubit $\psi = (\alpha, \beta)^T$ whose  corresponding density matrix is:
\begin{equation}
\scalemath{1.0}{
   \rho = \ket \psi \bra \psi =\begin{pmatrix}
                              \alpha\\
                               \beta
                               \end{pmatrix} (\alpha\ \beta) = \begin{pmatrix}
                                                                  |\alpha|^2 & \alpha\beta^*\\
                                                                   \alpha\beta^* & |\beta|^2
                                                                   \end{pmatrix}
}
\end{equation}
the mixture after applying COTP is $\scalemath{0.8}{\left\{\left( \frac{1}{2}, \ket \psi \right), \left( \dfrac{1}{2}, X \ket \psi \right) \right\}}$, and the density matrix is:
\begin{equation}
\scalemath{1.0}{
   \rho^\prime = \frac{1}{2} \rho + \frac{1}{2} X \rho X = \frac{1}{2} \begin{pmatrix}
                                                                                                          1 & \alpha\beta^* + \alpha^*\beta\\
                                                                                                           \alpha\beta^* + \alpha^*\beta & 1
                                                                                                           \end{pmatrix}
}
\end{equation}

This is not a maximally mixed state and the output probability depends on the measurement axis. When measuring along the $Z$ axis, whose basis states are $\ket 0 = (1, 0)^T$ and $\ket 1 = (0, 1)^T$, the probability of measuring $\ket 0$ and $\ket 1$ is the same, i.e. $P(\ket 0) = \bra 0 \rho^\prime \ket 0 = P(\ket 1) = \bra 1 \rho^\prime \ket 1 = \frac{1}{2}$, then there is no information about the initial states can be acquired from the measurement results. However, such a deduction is not held under some other measurement axes. The insecurity of this scheme can be proved by computing the probability of measuring an arbitrary state $\ket n = (x, y)^T, |x|^2 + |y|^2 = 1$:
\begin{equation}
\scalemath{0.95}{
   P(\ket n) = \frac{1}{2} \bra n \rho \ket n = \frac{1}{2} \left[1 + (x y^* + x^* y)(\alpha \beta^* + \alpha^* \beta) \right]
}
\end{equation}
According to this equation, the probability of measuring some states is not $\frac{1}{2}$. For instance, if the measurement is performed along $X$-axis whose basis states are $\ket + = \frac{1}{\sqrt{2}} (1, 1)^T$ and $\ket - = \frac{1}{\sqrt{2}} (1, -1)^T$, then $P(\ket +) = \bra + \rho^\prime \ket + = \frac{1}{2} (1 + \sin\theta \cos\phi)$ and $P(\ket -) = \bra - \rho^\prime \ket - = \frac{1}{2} (1 - \sin\theta \cos\phi)$, which depends on the initial states. Based on the measurement probability distribution, additional information about initial states is leaked.

\subsection{Quantum One-Time Pad (QOTP)}
\label{sec:qotp}

The insecurity of COTP in the quantum world can be fixed by introducing one more gate into the picture. Besides $k_1$ used to control whether to apply the Pauli-$X$ gate, Alice and Bob can also share one more bit $k_2$ to specify if a following Pauli-$Z$ gate will be performed. Alice then sends $\ket{\psi^\prime} = Z^{k_2} X^{k_1} \ket \psi$, and Bob can recover the state with $\ket \psi = X^{k_1} Z^{k_2} \ket{\psi^\prime}$. This scheme is called quantum one-time pad, or QOTP for short.
The mixture of QOTP
$\scalemath{0.8}{\left\{\left( \frac{1}{4}, \ket \psi \right), \left( \dfrac{1}{4}, X \ket \psi \right), \left( \dfrac{1}{4}, Z \ket \psi \right), \left( \dfrac{1}{4}, ZX \ket \psi \right) \right\}}$, or:
\begin{equation}
\scalemath{1.0}{
   \rho^\prime = \frac{1}{4} \rho + \frac{1}{4} X \rho X + \frac{1}{4} Z \rho Z + \frac{1}{4} ZX \rho XZ = \frac{1}{2} I
}
\label{eq:qopt_state}
\end{equation}

This is a maximally mixed state so the probability of measuring any state is $P(\ket n) = \frac{1}{2}$. Therefore, no information on the initial state can be learned with the measurement performed after QOTP was applied.

\subsection{Quantum Channels, Noise, and Errors} 

There are more general operations in quantum computing that cannot be expressed as quantum gates and we use the concept of \emph{quantum channels} to describe these. 
We can describe a quantum channel acting on a state $\rho$ via its Kraus representation  
as:
\begin{equation}
\scalemath{1.0}{
    \mathcal{E}(\rho) = \sum_i K_i \rho K_i^\dagger
}
\label{eq:quantum_channel}
\end{equation}
where $K_i$ are called Kraus operators satisfying ${\sum_i K_i^\dagger K_i = I}$.  
There are other representations, such as the Choi-matrix representation, we refer readers to~\cite{wood2015tensor} for more details. 

As a general approach, quantum channels can also be used to model the noisy process in quantum computing. 
Noise in quantum computing arises from various sources, including temperature fluctuations, electromagnetic interference, and imperfections in hardware components. 
These factors collectively introduce errors that can distort quantum operations. 
To be more specific in terms of noise sources, errors can be classified as thermal relaxation errors, measurement errors, Pauli errors, and so on. 
We discuss how to model these errors and deduce the theoretical formula in more detail in Section~\ref{sec:noise_and_errors}.

\subsection{Workflow of Cloud Quantum Computing}

All mathematical computations are used to model quantum circuits at the logic level. With quantum software development kits, such as Qiskit~\cite{Qiskit}, mathematical descriptions can be implemented as quantum circuits. Quantum circuits need to be further processed, the process is referred to as transpiling, to be transformed into instructions that can be executed on a specific quantum computer satisfying its requirements.

One quantum circuit typically needs to run numerous times in order to obtain the statistical result due to the probabilistic nature of quantum algorithms. One execution within a quantum circuit is often called one shot. This shot-by-shot execution enables the gathering of data and investigation of potential consequences. A key component of shot-based quantum computing is the reset operation. The qubits are reset between executions, usually $\ket 0 = (1,0)^T$, which makes sure that each succeeding shot starts from a specified state.

In quantum computers provided by cloud platforms, users submit multiple-shot quantum circuits to the quantum hardware, and these tasks are carried out, with each shot denoting a distinct computation. Each shot may be followed by measurements that reveal important details about the behavior and statistical characteristics of the quantum system. 
After all shots have been completed, the user will receive the final results.

\section{Threat Model}
\label{sec:threat_model}

\begin{figure}
    \includegraphics[width=0.8\linewidth]{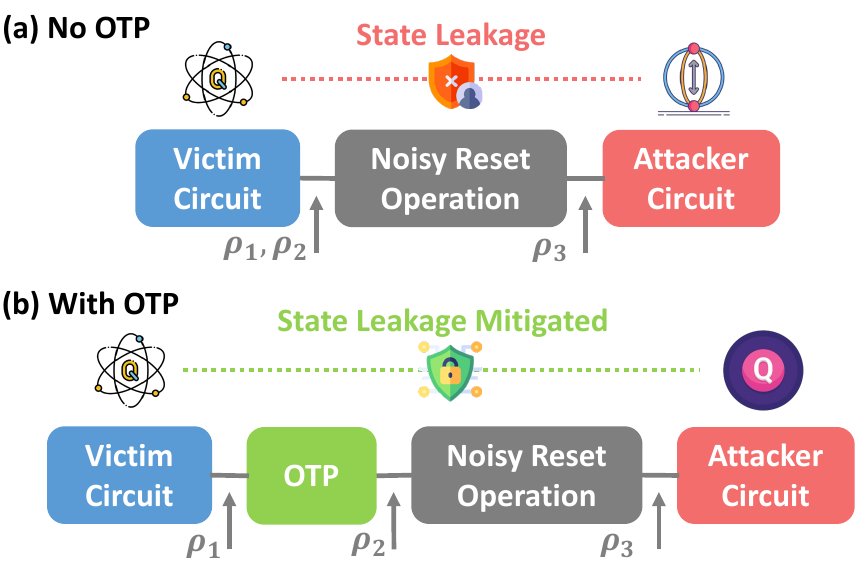}
    \centering
    \caption{Schematic of the threat model. (a) Without OTP, the state of the victim circuit is leaked to the attacker circuit; (b) with OTP, the state leakage can be mitigated.}
    \label{fig:threat_model}
\end{figure}

We first assume a strong attacker in order to later provide reliable assurances on the security of our defense. We assume that quantum computers can be shared, enabling the alternate execution of circuits from different users on a group of qubits on a quantum computer. We assume that between the shots of circuits, there is a reset operation. Several reset operations can be performed and are analyzed in this work.

We assume that there are two different types of users: victim users and attacker users. This work demonstrates OTP can be used after victim execution, but before reset operation execution, to prevent information leakage to the attacker user, as shown in Figure~\ref{fig:threat_model}. We assume that the attacker user has a reliable means to alternate execution with the victim user on the same qubits to collect measurement data that he or she uses to try to leak the state information from the victim.

We assume that when the victim finishes their computation and reads out their qubits, the attacker wants to learn the outcomes of these quantum programs. We assume that the victim and the attacker will run their programs consecutively for a sufficient number of times, enabling the attacker to gather statistical data from their respective applications. We assume, in particular, that the owner of the quantum computer has strong logical isolation such that the attacker cannot directly access the victim's outputs. If not, it would be simple to determine the victim's computation results, negating the necessity for side channels and information leaking analysis.

\section{One-Time Pad in Quantum Computing}

In this section, we analyze how the one-time pad can be applied to mitigate state leakage in quantum computing jobs. For simplicity, the deduction is based on single-qubit state, while it can be extended to multi-qubit states.

In our assumed setting, shown in Figure~\ref{fig:threat_model},
the victims finish executing one shot of their circuit, and then the system reset mechanism is triggered to reset the qubit. Typically, this can be a reset instruction, or simply idle the system for a long time to let qubits decohere to the $\ket{0}$ state. 
Finally, the attacker measures the state leakage.

The existence of state leakage is due to the noise and errors in reset operations. If the reset operation can completely reset states, then no information will persist into the following execution. However, if one scheme before the reset operation can change all states into one same state (such as QOTP that we demonstrated in Section~\ref{sec:qotp}), or states that the reset operation can further change into one same state (such as COTP with the reset instruction that we will discuss in the following), then it can mitigate state leakage. As we introduced in Section~\ref{sec:qotp}, QOTP transforms any state into the maximally mixed state and thus can be utilized to eliminate state leakage. 
This section mainly discusses the case for COTP. Later, we will show the requirements under which COTP is able to minimize state leakage.

\begin{figure}[t]
     \centering
     \begin{subfigure}[b]{0.5\linewidth}
         \centering
         \includegraphics[width=0.90\textwidth]{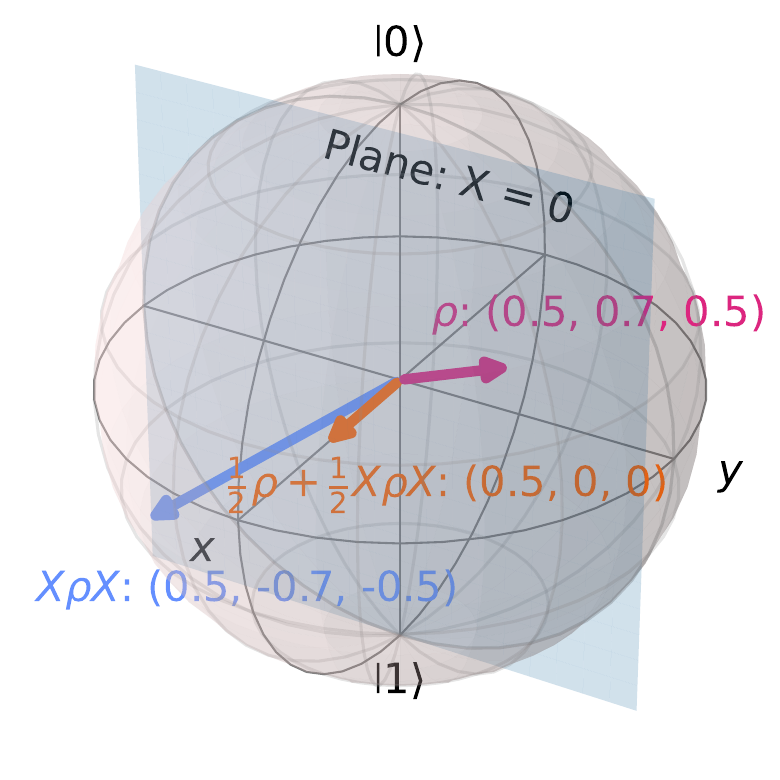}
         \caption{\small COTP}
         \label{fig:bloch_cotp}
     \end{subfigure}~
     \begin{subfigure}[b]{0.5\linewidth}
         \centering
         \includegraphics[width=0.90\textwidth]{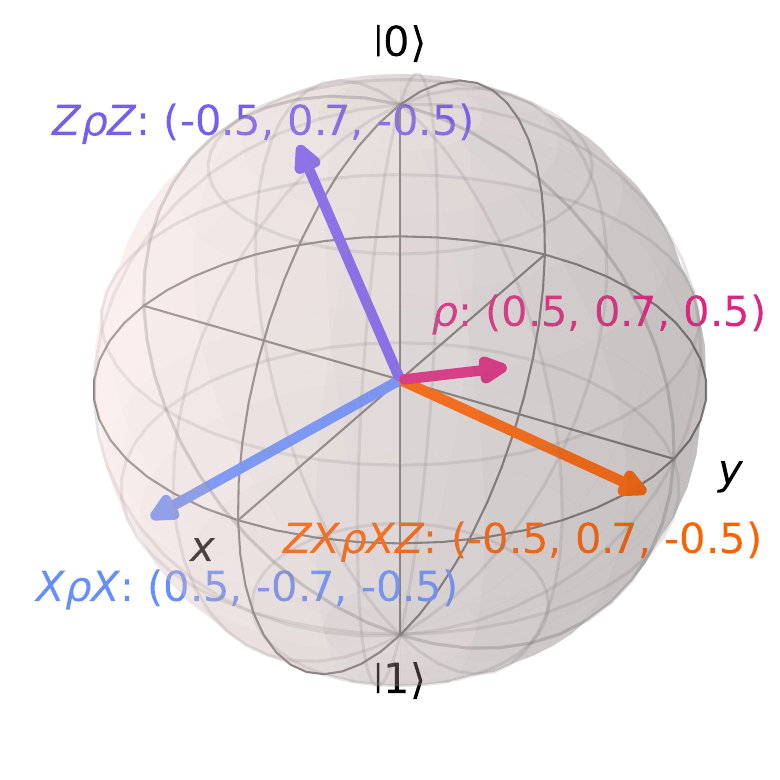}
         \caption{\small QOTP}
         \label{fig:bloch_qotp}
     \end{subfigure}
        \caption{\small Schematic of how states are transformed with OTP. The pink vector shows an arbitrary state, while others show the state after applying gates of OTP. (a) With COTP, the final states (the orange vector) are along the axis of the generalized Pauli-$X$ gate; (b) with QOTP, the final states (no show in the figure) are the original points.}
        \label{fig:bloch_otp}
\end{figure}

\subsection{States of Victim Circuits}

A density matrix should be used to represent a general case of the states after the victim circuit finishes. Here $\rho_1$ is the state after the victim finishes, as shown in Figure~\ref{fig:threat_model}b.
\begin{equation}
    \rho_1(\Vec{r}) = \frac{1}{2}(I + \Vec{r}\cdot \Vec{\sigma})
        =  \frac{1}{2} \begin{pmatrix}
                1 + r\cos\theta & re^{-i\phi}\sin{\theta}\\
                re^{i\phi}\sin\theta & 1 - r\cos{\theta}
            \end{pmatrix}
    \label{eq:rho1}
\end{equation}
where $I$ is the identity matrix, $\Vec{\sigma} = (\sigma_x, \sigma_y, \sigma_z)$ is the vector of three Pauli matrices, and $\Vec{r} = r(\sin\theta\cos\phi, \sin\theta\sin\phi, \cos\theta)$ is the vector in Bloch sphere to represent the states. For one state, $r = |\Vec{r}| \leq 1$, i.e. the vector is encircled inside the Bloch sphere shown in Figure~\ref{fig:bloch_otp}.

One distinction of applying OTP to quantum computing from other scenarios such as quantum teleportation is that at the end of circuits, there are usually measurements to obtain the computing results. Measurement is not a unitary operation and will collapse states. 
For example, when measuring along the $Z$ axis whose eigenstates are $\ket 0$ and $\ket 1$, without knowing the measurement results, the state after the measurement can be represented as $\rho = p \ket 0 \bra 0 + (1-p) \ket 1 \bra 1$, where $p$ is the probability of measuring $\ket 0$.  Equation~\ref{eq:rho1} already captures any state, and the correspondence is $r = 2p - 1$, i.e. if the eigenstates of the measurement is $\Vec{n}$ and $\Vec{-n}$ where $\Vec{r} = (2p - 1)\Vec{n}$, then $\rho_1(p) = p\ket n \bra n + (1-p) \ket{-n}\bra{-n} = \rho_1(\Vec{r})$.

\subsection{States After OTP}
\label{sec:rho2}

For a general study, assume the gate used by COTP is a generalized Pauli-$X$ gate that has eigenstate $\ket n$ with eigenvalue 1 and eigenstate $\ket{-n}$ with eigenvalue -1, where $\Vec{n} = (\sin\theta_n\cos\phi_n, \sin\theta_n\sin\phi_n, \cos\theta_n)$, so:
\begin{equation}
\scalemath{0.9}{
    X_{\Vec{n}} = \ket n \bra n - \ket{-n} \bra{-n} = \begin{pmatrix}
               \cos\theta_n & e^{-i\phi_n}\sin\theta_n\\
                e^{i\phi_n}\sin\theta_n & -\cos\theta_n
            \end{pmatrix}}
\end{equation}
for a special case when $\theta_n = \frac{\pi}{2}$ and $\phi_n = 0$, $X_{\Vec{n}} = X$.

The state after COTP,
shown as $\rho_2$ in Figure~\ref{fig:threat_model}b,
is:
\begin{equation}
\scalemath{0.9}{
\begin{aligned}
    &\rho_2(\Vec{r}, \Vec{n})  = \frac{1}{2}\rho_1(\Vec{r}) + \frac{1}{2} X_{\Vec{n}} \rho_1(\Vec{r}) X^\dagger_{\Vec{n}}\\
        & = \frac{1}{4}\begin{pmatrix}
               &2 + 2r\cos\theta\cos^2\theta_n + re^{-i(\phi + \phi_n)}(e^{2i\phi} + e^{2i\phi_n})\cdot\\
               &\sin\theta\sin\theta_n\cos\theta_n,\\
               & r\sin\theta_n\{2e^{-i\phi_n}\cos\theta\cos\theta_n + \\
               & e^{-i\phi}[1+e^{2i(\phi - \phi_n)}]\sin\theta\sin\theta_n\},\\
               & r\sin\theta_n\{2e^{i\phi_n}\cos\theta\cos\theta_n + \\
               & e^{i\phi}[1+e^{2i(\phi_n - \phi)}]\sin\theta\sin\theta_n\},\\
               &2 - 2r\cos\theta\cos^2\theta_n - re^{-i(\phi + \phi_n)}(e^{2i\phi} + e^{2i\phi_n})\cdot\\
               &\sin\theta\sin\theta_n\cos\theta_n
            \end{pmatrix}
\end{aligned}
}
\label{eq:rho_2}
\end{equation}

When $X_{\Vec{n}} = X$, the state above is simplified as:
\begin{equation}
\scalemath{0.9}{
    \rho_{2, X}(\Vec{r}) = \frac{1}{2}\begin{pmatrix}
                1 & r\sin\theta\cos\phi \\
                r\sin\theta\cos\phi & 1
            \end{pmatrix}
    \label{eq:rho2_x}
}
\end{equation}

On the other hand, after QOTP, the state is $\rho_2(\Vec{r}) = \frac{1}{2} I$ as shown in Equation~\ref{eq:qopt_state}. This holds for a generalized QOTP. The idea can be seen from Figure~\ref{fig:bloch_qotp}. Given the axes of the two generalized gates used in QOTP are orthogonal to each other, the state will be rotated along two axes independently to generate the mixture, and the vector of the mixture in the Bloch sphere is the original point.

\subsection{States After Reset Operations}
\label{sec:state_reset}

The states after the reset operation depend on the various implementations of the reset operation, such as the reset instruction, or simply idle qubits to decohere. Currently, the reset instruction is typically implemented as one mid-circuit measurement and one Pauli-$X$ gate conditioned on the measurement result. If the measurement result is $\ket 0$, then it is already in the ground state so the Pauli-$X$ gate will not be applied. Otherwise, the Pauli-$X$ gate will be applied to flip the state to $\ket 0$.

For a generalized reset instruction, suppose the measurement is along the axis $\Vec{m} = (\sin\theta_m\cos\phi_m, \sin\theta_m\sin\phi_m, \cos\theta_m)$, or correspondingly $\ket m = \cos\frac{\theta_m}{2} \ket 0 + e^{i\phi_m} \sin\frac{\theta_m}{2} \ket 1$. Without losing generality, we choose to start from Equation~\ref{eq:rho2_x}. The possibility of measuring $\ket m$ and $\ket{-m}$ with $\rho_{2, X}(\Vec{r})$ is:
\begin{equation}
\scalemath{0.85}{
\begin{aligned}
    & P_{X}(\ket{m}|\ \Vec{r}) = \bra{m} \rho_{2, X}(\Vec{r}) \ket{m} = \frac{1}{2} ( 1 + r \sin\theta\cos\phi \sin\theta_m\cos\phi_m )\\
    & P_{X}(\ket{-m}|\ \Vec{r}) = \bra{-m} \rho_{2, X}(\Vec{r}) \ket{-m} = \frac{1}{2} ( 1 - r \sin\theta\cos\phi \sin\theta_m\cos\phi_m )
\end{aligned}
}
\label{eq:prob_reset}
\end{equation}
notice that after determining the axis of the measurement in the reset instruction, the conditional gate is determined, i.e. if the ground state is chosen to be $\ket m$, then the conditional gate must change $\ket{-m}$ to $\ket m$, i.e., it functions the same as $X_{\Vec{m}}$.

For other types of implementations of the reset operation, the process can be modeled with the quantum channel $\mathcal{E}$ in Equation~\ref{eq:quantum_channel}, since the reset operation is not unitary.

Besides, as we explained, the state leakage exists due to the noise and errors. For the reset instruction, the measurement error plays a role when measuring $\ket{-m}$ but reporting to measure $\ket{m}$, it keeps $\ket{-m}$ unchanged, or the opposite. The gate error works when $\ket{-m}$ will not be correctly rotated to $\ket m$. For a generalized reset operation with errors, the state after the reset operation,
shown as $\rho_3$ in Figure~\ref{fig:threat_model}b,
can be represented as:
\begin{equation}
\scalemath{0.9}{
    \rho_3 (\Vec{r}) = \mathcal{E} (\rho_2) = \left( 1 - \sum_{i = 1}^n p_i(\Vec{r}) \right) \ket m \bra m + \sum_{i = 1}^n p_i(\Vec{r}) \ket{e_i(\Vec{r})}\bra{e_i(\Vec{r})}}
    \label{eq:rho_4}
\end{equation}
where $\ket{e_i(\Vec{r})}\bra{e_i(\Vec{r})}$ is one state to which the noise and errors cause, and $p_i(\Vec{r})$ is the probability of this result.

\subsection{Measurement of the State Leakage}

The state leakage can be measured with a subsequent measurement. If $p_i$ is independent of the initial state $\Vec{r}$, or if the $p_i$ is dependent on the initial state $\Vec{r}$ but the following reset operation removes this dependence, then no information will be leaked from the prior execution to the next. Otherwise, such dependence may be measured in attacker circuits, and then this state leakage can potentially lead to a bias in the computing results, or information leakage to the following execution.

\subsection{Multi-Qubit Case} 

For multi-qubit states that are not entangled, the extension of the previous discussion is straightforward since each qubit is independent. For a general multi-qubit state, the case is similar to the single-qubit case. As an example, consider a general 2-qubit state $\rho = \{a_{ij}\}$. After applying COTP independently on each qubit, the diagonal elements in the density matrix are both $\frac{1}{4}$, which means the probability of all cases are $\frac{1}{4}$ when measuring along the $Z$ axis, while leaving the off-diagonal elements to be a quarter of the summations of permutations between rows and columns. Because off-diagonal elements may be non-zero, they may be measured by attackers to retrieve the information of victims. In contrast, QOTP will evolve the state to be $\frac{1}{4} I_4$, which is still a maximally-mixed state. Whether multi-qubit entangled states can have more interesting behaviors will leave as a future work. For instance, it may be possible to measure one qubit to get information about other qubits due to entanglement, e.g., to ``phase kickback" in quantum algorithms~\cite{Lee_2016}.

\section{Noise and Errors}
\label{sec:noise_and_errors}

Noise and errors in the reset operation are the main reason for the state leakage. The noise and errors depend on the implementation of the reset operation. In this section, three types of reset operations will be discussed:
\begin{enumerate}
    \item {\bf Thermal Relaxation}: This reset approach simply idles the qubit for a long time to allow decoherence to occur. Currently, quantum computers on most cloud platforms, such as IBM Quantum, are mainly using this approach.
    \item {\bf reset instruction}: The typical implementation of a ``reset instruction'' is based on a mid-circuit measurement followed by a conditional Pauli-$X$ gate as introduced in Section~\ref{sec:state_reset}. IBM Quantum adopts this implementation.
    \item {\bf Measurement-less reset instruction}: For a theoretical study, we propose an imaginary reset instruction that is assumed to reset states but may maintain the state with a small probability. We will evaluate this theoretical reset instruction in Section~\ref{sec:eval_noise}.
\end{enumerate}

In the following, we only consider applying the quantum channel on $\rho_1(\Vec{r})$ (Equation~\ref{eq:rho1}, without OTP) and $\rho_{2, X}(\Vec{r})$ (Equation~\ref{eq:rho2_x}, with COTP of Pauli-$X$ gate), since QOTP evolve states to the maximally mixed state. The discussion of $\rho_{2, X}(\Vec{r})$ can be directly applied to COTP with the generalized Pauli-$X$ gate. The noise and errors in the gate used by OTP are not considered, and their influence on state leakage can be future work. For simplicity, we consider the axis of the attacker measurement to be along the $Z$ and $X$ axes. 

\subsection{Thermal Relaxation}
\label{sec:thermal}

In practice, qubits are constantly interacting with the environment, and through this process, quantum coherence is lost~\cite{zeh1970interpretation}. This process is called quantum decoherence, and is usually described by $T_1$ time, or the relaxation time, and $T_2$, or the dephasing time~\cite{10.1063/1.5089550}. For qubit in state $\ket 1$, the probability of measuring it to be $\ket 1$ after time $t$ is given by: $P(\ket 1) = e^{-\frac{t}{T_1}}$, where $T_1$ quantifies how the qubit decays to $\ket 0$. On the other hand, $T_2$ both describes the energy and phase loss, with $\frac{1}{T_2} = \frac{1}{2T_1} + \frac{1}{T_\phi}$, where $T_\phi$ is the pure dephasing time. According to this formula, it requires $T_2 \leq 2T_1$.

The thermal relaxation error channel can described in the Choi-matrix representation:
\begin{equation}
\scalemath{0.72}{
\begin{aligned}
    \Lambda & = \sum_{i, j} \ket i \bra j \otimes \mathcal{E}(\ket i \bra j)\\
    & = \begin{pmatrix}
        1 - p_1 (1-e^{-\gamma_1}) & 0 & 0 & e^{-\gamma_2} \\
        0 & p_1 (1-e^{-\gamma_1}) & 0 & 0 \\
        0 & 0 & p_0 (1-e^{-\gamma_1}) & 0 \\
        e^{-\gamma_2} & 0 & 0 & 1 - p_0 (1-e^{-\gamma_1})
    \end{pmatrix}
\end{aligned}
}
\end{equation}
where $\gamma_1 = \frac{T_1}{t}$ and $\gamma_2 = \frac{T_2}{t}$ is the ratio of the decoherence time to the time idled, and $p_0$ and $p_1$ are the populations of $\ket 0$ and $\ket 1$ at equilibrium, which is approximately $p_0 = 1$ and $p_1 = 0$ for most quantum computers.

Under this quantum channel, the state will be:
\begin{equation}
    \mathcal{E}(\rho) = \Tr_1\left[ \Lambda (\rho^T \otimes \mathbb{I}) \right]
\end{equation}
where $\Tr_1$ is the partial trace over subsystem 1. We refer readers to~\cite{wood2015tensor} for more details of the Choi-matrix representation.

Therefore, the state after thermal relaxation will be:
\begin{equation}
\scalemath{0.9}{
    \mathcal{E} \left[ \rho_1(\Vec{r}) \right] = \frac{1}{2} \begin{pmatrix}
        2 - e^{-\gamma_1} (1 - r\cos\theta) & e^{-\gamma_2-i\phi} r \sin\theta \\
        e^{-\gamma_2+i\phi} r \sin\theta & e^{-\gamma_1} (1 - r\cos\theta)
    \end{pmatrix}}
\end{equation}
\begin{equation}
\scalemath{0.9}{
    \mathcal{E} \left[ \rho_{2, X}(\Vec{r}) \right] = \frac{1}{2} \begin{pmatrix}
        2 - e^{-\gamma_1} & e^{-\gamma_2} r \sin\theta \cos\phi \\
        e^{-\gamma_2} r \sin\theta \cos\phi & e^{-\gamma_1}
    \end{pmatrix}}
\end{equation}

When the axis of the attacker measurement is along the $Z$ axis, the probability of measuring $-1$ is:
\begin{equation}
    P \left( -1 | \mathcal{E} \left[ \rho_1(\Vec{r}) \right] \right) = \frac{1}{2} e^{-\gamma_1} (1 - r \cos\theta)
\end{equation}
\begin{equation}
    P \left( -1 | \mathcal{E} \left[ \rho_{2, X}(\Vec{r}) \right] \right) = \frac{1}{2} e^{-\gamma_1 }
\end{equation}
Without COTP, the probability depends on the victim probability (recall $r = 2p(+1) - 1$) and its orientation, and thus the attacker can retrieve such information, while with COTP, the probability only depends on the decoherence time.

When the measurement axis is along the $X$ axis, the probability of measuring $-1$ is:
\begin{equation}
    P \left( -1 | \mathcal{E} \left[ \rho_1(\Vec{r}) \right] \right) = \frac{1}{2} (1 - e^{-\gamma_2} r \sin\theta \cos\phi)
\end{equation}
\begin{equation}
    P \left( -1 | \mathcal{E} \left[ \rho_{2, X}(\Vec{r}) \right] \right) = \frac{1}{2} (1 - e^{-\gamma_2} r \sin\theta \cos\phi)
\end{equation}
They are the same and depend on victim states. Thus, COTP cannot mitigate state leakage on this occasion.

\subsection{reset instruction}
\label{sec:reset}

For the reset instruction, the first part is the mid-circuit measurement, whose reported results are influenced by $M_{01}$ and $M_{10}$, which is the measurement error of preparing $\ket 1$ and measuring $\ket 0$ and preparing $\ket 0$ and measuring $\ket 1$ respectively. The state after the measurement can be represented as:
\begin{equation}
\scalemath{0.9}{
\begin{aligned}
       \mathcal{E}(\rho) = & \left[ \bra 0 \rho \ket 0 (1 - M_{10}) + \bra 1 \rho \ket 1 ( 1 - M_{01}) \right] \ket 0 \bra 0 + \\
       & (\bra 0 \rho \ket 0 M_{10} + \bra 1 \rho \ket 1  M_{01}) \ket 1 \bra 1
\end{aligned}}
\end{equation}

The second part is the conditional Pauli-$X$ gate, which is affected by the errors in measurement and also its own errors. There can be many types of errors for it, such as the bit-flip error, depolarizing error, etc. As an example, if we only assume the bit-flip error with the probability $p_{bf}$, then the state after the conditional Pauli-$X$ gate is:
\begin{equation}
\scalemath{0.8}{
\begin{aligned}
       \mathcal{E}(\rho) = & \left\{ \bra 0 \rho \ket 0 [(1 - M_{10}) + M_{10}p_{bf}] + \bra 1 \rho \ket 1 ( 1 - M_{01})(1 - p_{bf}) \right\} \ket 0 \bra 0 \\
       & + \left\{\bra 0 \rho \ket 0 M_{10}(1 - p_{bf}) + \bra 1 \rho \ket 1  [M_{01} + ( 1 - M_{01}) p _{bf}] \right\} \ket 1 \bra 1
\end{aligned}}
\end{equation}

Based on this, when the measurement axis is along the $Z$ axis, the probability of measuring $-1$ is:
\begin{equation}
\begin{aligned}
        P \left( -1 | \mathcal{E} \left[ \rho_1(\Vec{r}) \right] \right) & = \frac{1}{2} \{ [(M_{10} + M_{01}) (1 - p_{bf}) + p_{bf}] + \\ 
        & [(M_{10} - M_{01}) (1 - p_{bf}) - p_{bf}] r\cos\theta \}
\end{aligned}
\label{eq:reset_cotp_z}
\end{equation}
\begin{equation}
    P \left( -1 | \mathcal{E} \left[ \rho_{2, X}(\Vec{r}) \right] \right) = \frac{1}{2} [(M_{10} + M_{01}) (1 - p_{bf}) + p_{bf}]
\end{equation}

When the measurement axis is along the $X$ axis, the probability of measuring $-1$ is:
\begin{equation}
    P \left( -1 | \mathcal{E} \left[ \rho_1(\Vec{r}) \right] \right) = \frac{1}{2}
\end{equation}
\begin{equation}
    P \left( -1 | \mathcal{E} \left[ \rho_{2, X}(\Vec{r}) \right] \right) = \frac{1}{2}
\end{equation}

COTP masks the dependence on the input parameter thus mitigating the state leakage in both axes, and there is no state leakage even for no OTP when measuring along the $X$ axis. In addition, as Equation~\ref{eq:reset_cotp_z} shows, generally $M_{10}$, $M_{01}$, and $p_{bf}$ are small, under which case the direction of the state leakage pattern depends on $M_{10} - M_{01}$. Usually, $M_{10} < M_{01}$ due to the decoherence in the measurement process, and thus the pattern will be similar to other reset operations. However, sometimes $M_{10} > M_{01}$. This leads to a reverse direction of the state leakage pattern, which is also shown in~\cite{10.1145/3548606.3559380}.

\subsection{Measurement-less reset instruction}
\label{sec:imag_reset}

Lastly, we consider a theoretical reset instruction, which is a simplified reset instruction, which either leaves the state unchanged with the probability $p_r$ or resets the state with the probability $1 - p_r$:
\begin{equation}
    \mathcal{E}(\rho) = p_r \rho + (1 - p_r) \ket 0 \bra 0
\end{equation}

Therefore, the state after this reset will be:
\begin{equation}
\scalemath{0.9}{
    \mathcal{E} \left[ \rho_1(\Vec{r}) \right] = \frac{1}{2}\begin{pmatrix}
                2 - p_r (1 - r\cos\theta) & p_r re^{-i\phi}\sin{\theta}\\
                p_r re^{i\phi}\sin\theta & p_r (1 - r\cos{\theta})
    \end{pmatrix}}
\end{equation}
\begin{equation}
\scalemath{0.9}{
    \mathcal{E} \left[ \rho_{2, X}(\Vec{r}) \right] = \frac{1}{2} \begin{pmatrix}
                2 - p_r & p_r r\sin\theta\cos\phi \\
                p_r r\sin\theta\cos\phi & p_r
    \end{pmatrix}}
\end{equation}

When the measurement axis is along the $Z$ axis, the probability of measuring $-1$ is:
\begin{equation}
    P \left( -1 | \mathcal{E} \left[ \rho_1(\Vec{r}) \right] \right) = \frac{1}{2} p_r (1 - r \cos \theta)
\end{equation}
\begin{equation}
    P \left( -1 | \mathcal{E} \left[ \rho_{2, X}(\Vec{r}) \right] \right) = \frac{1}{2} p_r
    \label{eq:reset_cotp_z}
\end{equation}

When the measurement axis is along the $X$ axis, the probability of measuring $-1$ is:
\begin{equation}
    P \left( -1 | \mathcal{E} \left[ \rho_1(\Vec{r}) \right] \right) = \frac{1}{2} (1 - p_r r \sin\theta \cos\phi)
\end{equation}
\begin{equation}
    P \left( -1 | \mathcal{E} \left[ \rho_{2, X}(\Vec{r}) \right] \right) = \frac{1}{2} (1 - p_r r \sin\theta \cos\phi)
    \label{eq:reset_cotp_x}
\end{equation}

Both are the same as the case of thermal relaxation if considering $p_r = e^{-\gamma_1}$ and $p_r = e^{-\gamma_2}$. This theoretical reset instruction can be considered as the special thermal relaxation process where the decoherence is isotropic.

\section{Requirements for Classical One-Time Pad}
\label{sec:requirements}

As discussed in the previous section, after applying COTP, the probability may include $r$ and $\theta$, which is related to the states of the previous execution. However, we will discuss in this section that with the correct design of quantum computer systems, this information cannot be effectively measured.

\subsection{Victim Circuit Measurement Axis}
\label{sec:req_meas}

Measuring along different axes is required in many quantum algorithms. This can be done by adding quantum gates before the measurement. To make the collapsed states correct, additional gates also need to be added after the measurement. However, gates after the measurement are optional in many cases, such as in the final measurement since the state is not concerned any further.

If all measurements are on the same axis and the gate after the measurement is not added, such as in Qiskit and IBM Cloud, where the measurement axis is along $Z$ axis, the COTP is enough since $\theta = 0$ and the off-diagonal elements in Equation~\ref{eq:rho2_x} is $0$. The reason is that the states after the measurement can only be one of the two eigenstates of the measurement ($\ket 0$ and $\ket 1$ in the case of $Z$ axis), and thus the case is totally the same as to encode the classical bits.

However, if the feature of changing the axis of the measurement is supported natively, then COTP cannot fully obliterate dependence on the input parameters as exemplified in the previous section. For the native support, the gate after the measurement is necessary, since the measurement may also be used in the middle of the circuit, and thus the state is needed to be one of the eigenstates of the measurement. 

In addition, in some cases, some qubits will not be measured at the end of the circuits, such as ancilla qubits. Consequently, COTP may not mitigate the state leakage in both cases.

\subsection{Reset Operation}

According to Equation~\ref{eq:rho2_x}, if the gate used in COTP is the Pauli-$X$ gate, the axis of the measurement in the reset instruction can be chosen to be along the $Z$ axis. Under this circumstance, COTP can get rid of the dependence on the according to Equation~\ref{eq:reset_cotp_z} and Equation~\ref{eq:reset_cotp_x}. Since the reset mechanism is supposed to be supported natively in quantum computer systems and cannot be tuned by users, this can be a direct solution to mitigate state leakage.

This is not the only solution. To make the state after COTP indistinguishable, the measurement axis can be any axis in the plane perpendicular to the axis of the gate used in OTP. The idea is shown in Figure~\ref{fig:bloch_cotp}. Because the gate used in COTP rotates the state around its axis for $\pi$, the component parallel to the axis is the same, while the component orthogonal to the axis is the opposite. The state after COTP is along its gate axis. The measurement is one projection operation to its eigenvectors, and thus for the measurement with any axis in the plane perpendicular to the axis of the gate used in OTP, the probability of measuring two results will be 0.5. Consequently, COTP only hides components orthogonal to the axis of its gate. 
Note, for a generalized reset operation in Equation~\ref{eq:rho_4}, such as the decoherence, COTP will not help in most cases.

\subsection{Requirement Summary}

In summary, due to that COTP can only hide information about the components orthogonal to the axis of its gate, and the measurement is a non-unitary operation that will only measure information corresponding to some axis, these two features intertwine with each other and lead to a simple design to mitigate state leakage: COPT with Pauli-$X$ gate + mid-measurement along $Z$ axis and Pauli-$X$ gate conditioned on the measurement results. The axis is not unique and can be changed based on the discussion in this section. This design is already able to be implemented in most cloud platforms. If such a reset instruction is not available, COTP can also be applied with small errors in operations, though cannot completely mitigate state leakage, which will be evaluated in Section~\ref{sec:eval}.

To conclude, if the quantum channel of the reset operation has some symmetries that cancel out the off-diagonal elements of $\rho_{2, X}$ (Equation~\ref{eq:rho2_x}), or get rid of the dependence on the input parameter $r$, $\theta$, and $\phi$ in $\rho_{2, X}$ or $\rho_{2}$ (Equation~\ref{eq:rho_2}), then COTP can be applied to mitigate state leakage.

\section{Evaluation of State Leakage}
\label{sec:eval}

This section presents the evaluation results on both the real quantum computer and the simulator.

\subsection{Experiment Setup}

The settings of quantum circuits are shown in Figure~\ref{fig:threat_model}. The state of the victim circuit is generated by a rotational $X$ gate with angle $\alpha$ chosen from nine vales of $\{0, \frac{1}{8} \pi, \frac{1}{4} \pi, \dots, \frac{7}{8} \pi, \pi\}$, and then followed by a Pauli-$Z$ gate, and finally evolves the state to $\ket \psi = \cos\frac{\alpha}{2} \ket 0 + \sin\frac{\alpha}{2} \ket 1$. In the end, the state is measured along the $Z$ or $X$ axis, which corresponds to set $\theta = 0, \phi = 0$ or $\theta = \frac{\pi}{2}, \phi = 0$ respectively in Equation~\ref{eq:rho1}. Note that $r = 2P(+1) - 1$ after the measurement, where $P(+1)$ is the probability of measuring $\ket 0$ when along the $Z$ axis and measuring $\ket +$ when along the $X$ axis. The measurement is assumed to also collapse the states to its eigenstates, so it will be followed by a Hadamard gate when measuring along the $X$ axis, as discussed in Section~\ref{sec:req_meas}. 
After the victim circuit is the reset mechanism. The first step is to randomly apply gates of COTP or QOTP, which are chosen to be the Pauli-$X$ gate and Pauli-$Z$ gate, or no gate is applied if no OTP is employed. Then one of the reset operations discussed in Section~\ref{sec:noise_and_errors} is applied to reset the state. For experiments on real quantum computers, only the default delay and the supported reset instruction are used, while the measurement-less reset instruction is evaluated in experiments on simulators. In the end, one measurement simulating the attacker's behavior measures the state leakage. This measurement will also be along the $Z$ or $X$ axis. For each parameter set, the experiments were done 10 times, with each experiment being performed 10,000 shots.

In Section~\ref{sec:eval_qc}, the state leakage results on the real quantum computer {\tt ibmq\_jakarta} are shown, which is a 7-qubit machine on IBM Quantum. In Section~\ref{sec:eval_sim} and Section~\ref{sec:eval_noise}, {\tt AerSimulator} provided in Qiskit with noise model will be used for testing state leakage with different reset operations and error rates. The simulator is used because only the simulator can be tuned with different error rates, and noise and errors are the same over time, while the noise and errors on real quantum computers are volatile.

The parameter space is infinite and thus must be limited for evaluations. For a general evaluation, many parameters can further be tested, such as the phase $\phi$, the measurement axis angles, and so on. Nevertheless, the general discussion was presented in previous sections, and evaluations in the following demonstrated the idea without losing generality.

\subsection{State Leakage in Real Quantum Computers}
\label{sec:eval_qc}

\begin{figure*}
     \centering
     \begin{subfigure}[b]{\textwidth}
         \centering
         \includegraphics[width=0.95\textwidth]{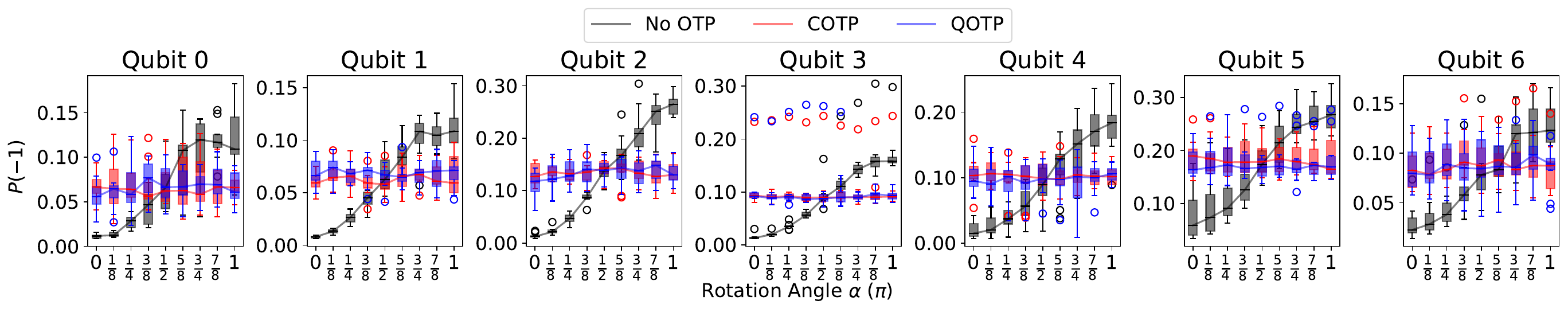}
         \caption{\small 250 ns delay and the measurement axis is along the $Z$ axis.}
         \label{fig:qc_delay_z}
     \end{subfigure}\\
     \begin{subfigure}[b]{\textwidth}
         \centering
         \includegraphics[width=0.95\textwidth]{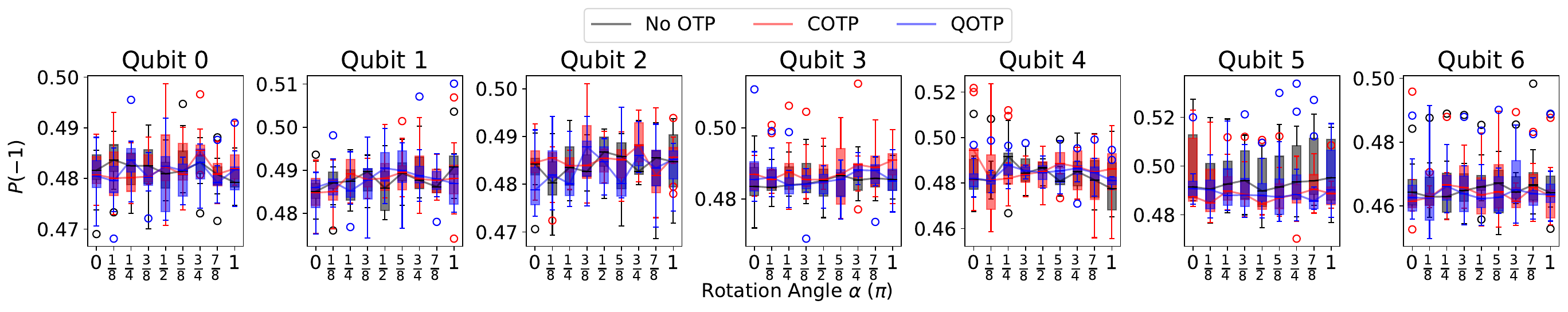}
         \caption{\small 250 ns delay and the measurement axis is along the $X$ axis.}
         \label{fig:qc_delay_x}
     \end{subfigure}
     \begin{subfigure}[b]{\textwidth}
         \centering
         \includegraphics[width=0.95\textwidth]{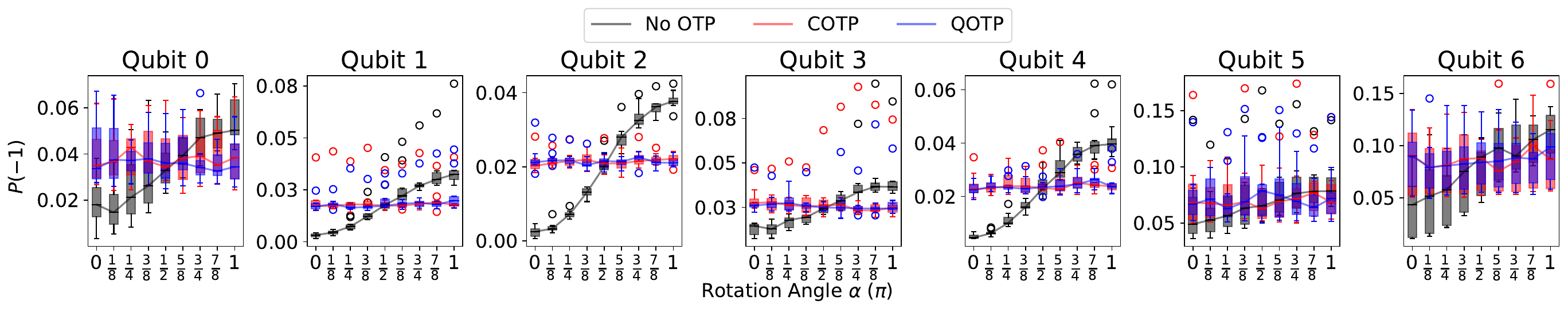}
         \caption{\small reset instruction and the measurement axis is along the $Z$ axis.}
         \label{fig:qc_reset_z}
     \end{subfigure}
     \begin{subfigure}[b]{\textwidth}
         \centering
         \includegraphics[width=0.95\textwidth]{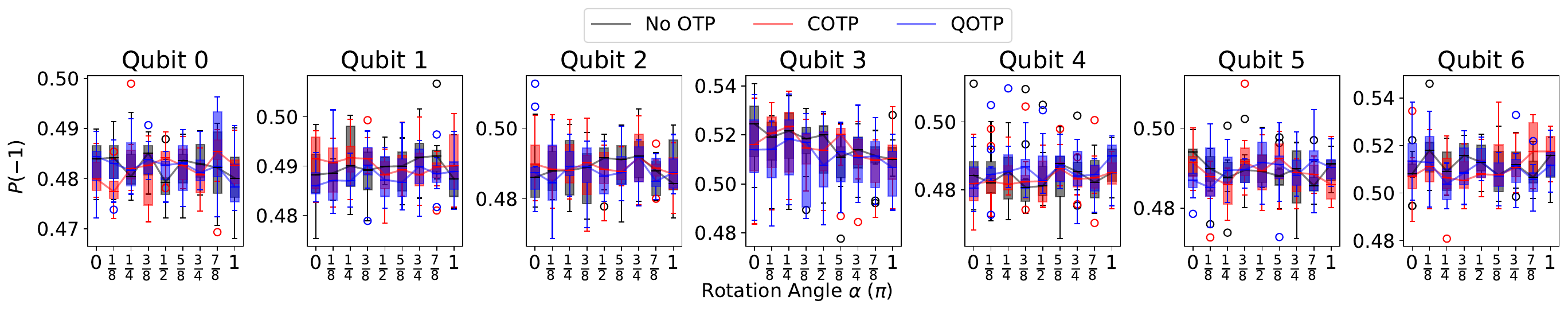}
         \caption{\small reset instruction and the measurement axis is along the $X$ axis.}
         \label{fig:qc_reset_x}
     \end{subfigure}
        \caption{\small $P(-1)$, the probability of attackers measuring $-1$ on {\tt ibmq\_jakarta} on IBM Quantum. The state leakage is shown in (a) and (c) without OTP (black lines) by the apparent dependence of $P(-1)$ on $\alpha$. (a) The reset operation is 250 ns delay (default value on IBM Quantum) and the measurement axis is the $Z$ axis; (b) the reset operation is 250 ns delay and the measurement is the $X$ axis; (c) the reset operation is 250 ns delay and the measurement axis is the $X$ axis. (c) the reset operation is the default reset instruction and the measurement axis is the $Z$ axis; (d) the reset operation is the default reset instruction and the measurement axis is the $X$ axis.}
        \label{fig:qc_result}
\end{figure*}

The state leakage on the real quantum computer {\tt ibmq\_jakarta} is shown in Figure~\ref{fig:qc_result}. In Figure~\ref{fig:qc_delay_z} and Figure~\ref{fig:qc_reset_z}, the black lines, which are the result without OTP, show that $P(-1)$ depends on the victim state parameter $\alpha$. Some qubits are less noisy and present a clear pattern, such as qubits 1-4, while the other qubits are more noisy and the standard deviation is large, which is due to the instability of noise and errors of quantum computers. Still, a pattern similar to the trigonometric function is shown. As discussed before, both COTP and QOTP can mitigate state leakage when the victim measures along the $Z$ axis, and this is proved by the flat lines in these figures.

However, in Figure~\ref{fig:qc_delay_x} and Figure~\ref{fig:qc_reset_x}, there is no clear dependence for all three cases of OTP when measuring along the $X$ axis, this is predicted in Section~\ref{sec:noise_and_errors} for the reset instruction but not the thermal relaxation. One reason is noise and errors from other sources, such as the gates in the victim circuits to prepare the states and the measurement of the attacker circuit. Another reason is that $T_2$ is very small on most qubits so that dependence is small. Usually, $T_1$ is much larger than $T_2$, e.g., on {\tt ibmq\_jakarta}, $T_1$ is usually between 100 ns and 200 ns, while $T_2$ is usually less than 100 ns. Therefore, according to Section~\ref{sec:thermal}, the dependence is much smaller when measuring along the $X$ axis than measuring along the $Z$ axis. In any case, OTP can suppress the state leakage.

\subsection{State Leakage with Different Reset Operations}
\label{sec:eval_sim}

\begin{figure}
     \centering
     \begin{subfigure}[b]{0.5\textwidth}
         \centering
         \includegraphics[width=0.728\textwidth]{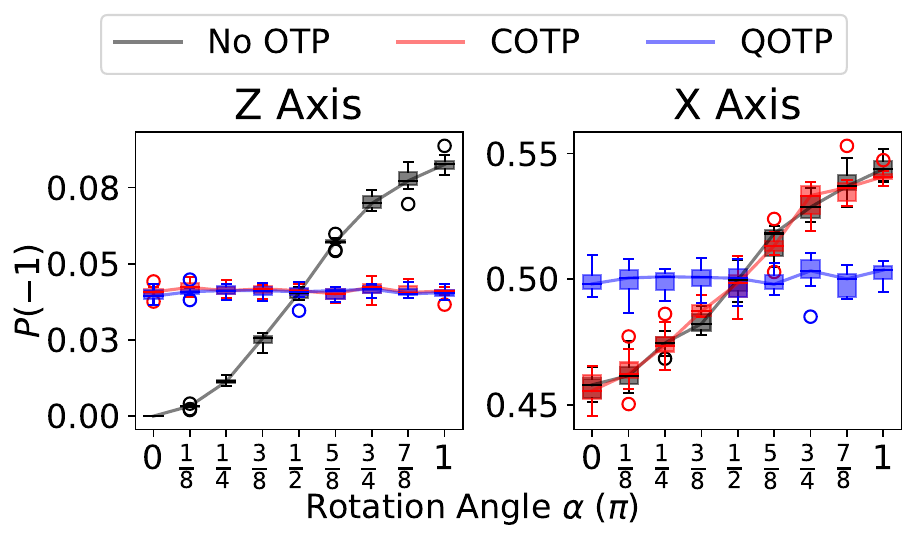}
         \caption{\small Thermal Relaxation (250 ns delay).}
         \label{fig:sim_delay}
     \end{subfigure}\\
     \begin{subfigure}[b]{0.5\textwidth}
         \centering
         \includegraphics[width=0.728\textwidth]{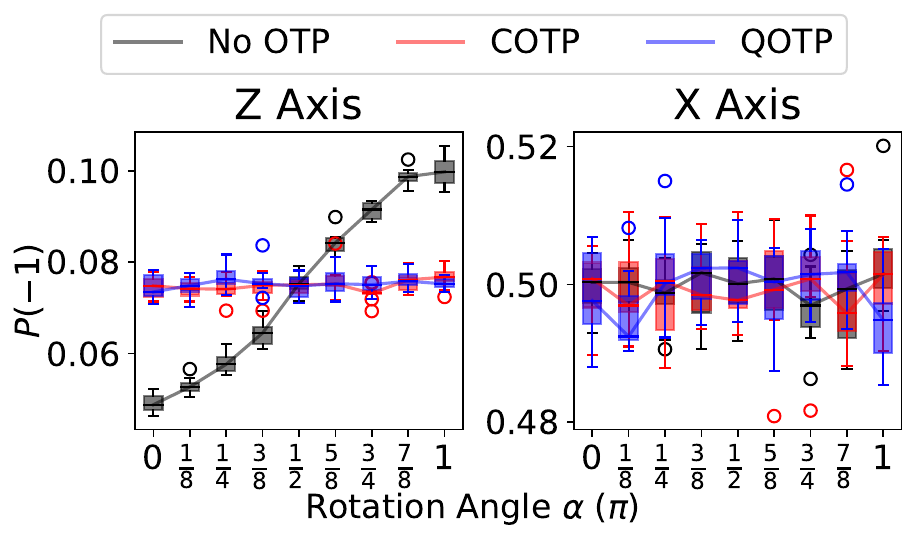}
         \caption{\small reset instruction.}
         \label{fig:sim_reset}
     \end{subfigure}
     \begin{subfigure}[b]{0.5\textwidth}
         \centering
         \includegraphics[width=0.728\textwidth]{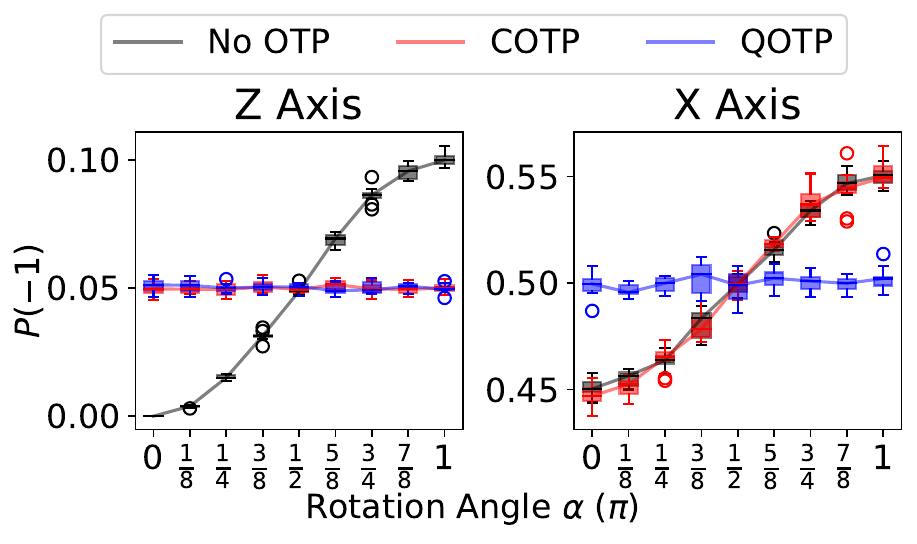}
         \caption{\small Measurement-less reset instruction.}
         \label{fig:sim_imag_reset}
     \end{subfigure}
        \caption{\small $P(-1)$, the probability of attackers measuring $-1$ on the simulator. (a) The reset operation is 250 ns delay (default value on IBM Quantum) and $T_1 = T_2 = 100$ ns; (b) the reset operation is the reset instruction and $M_{10} = 0.05, M_{01} = 0.10$ and no error on Pauli-$X$. (c) The reset operation is the measurement-less reset instruction and $p_r = 0.1$.}
        \label{fig:sim_result}
\end{figure}

The state leakage results of three reset operations on the simulator are shown in Figure~\ref{fig:sim_result}. For the delay, $T_1 = T_2 = 100$ ns. For the reset instruction, $M_{10} = 0.05, M_{01} = 0.10$. For the measurement-less reset instruction, $p_r = 0.1$. The results are consistent with the previous discussion in Section~\ref{sec:noise_and_errors}. All three operations will have considerable state leakage when measuring along the $Z$ axis and no OTP is applied. While QOTP can eliminate the dependence on $\alpha$ in all cases, COTP can only achieve this for the reset instruction and has a similar pattern as no OTP in the other two cases.

\subsection{Noise and Errors}
\label{sec:eval_noise}

To quantify the state leakage with different noise and error rates, we define signal-to-noise ratio (SNR), which is similar to SNR which is widely used to measure the signal in the background of noise. In this attack, the state leakage pattern can be approximately quantified with the measure below:
\begin{equation}
\scalemath{0.88}{
    SNR = \frac{mean_{exp}[P(-1|\alpha=\pi)] - mean_{exp}[P(-1|\alpha=0)]}{mean_\alpha[\sigma(\alpha) * \sqrt{\frac{n_{exp}}{n_{exp}-1}}]}
}
\end{equation}
where $P(-1|\alpha = x)$ is the probability of measuring $-1$ when $\alpha = x$, and the mean value of it is over all the experiments. $\sigma(\alpha)$ is the standard deviation of all the experiments given $\alpha$. The factor $\sqrt{\frac{n_{exp}}{n_{exp}-1}}$ is Bessel's correction to estimate the unbiased standard deviation, where $n_{exp}$ is the number of experiments. This quantity describes the degree of state leakage, or in the view of security and privacy, how capable attackers can retrieve the input state from the results. The larger means the state leakage is more remarkable, or attackers can retrieve the victim's information more easily.

Given $P(-1|\alpha = x)$, SNR can be computed following the formula in Section~\ref{sec:noise_and_errors}. Because the measurement results can only be $+1$ or $-1$, it follows the Bernoulli distribution. Thus, for one experiment consisting of $n$ shots, the expectation value is $P(-1|\alpha = x)$. Assuming the independence among each shot, the standard deviation is $\sqrt{P(-1|\alpha = x)*[1-P(-1|\alpha = x)] / n}$. So the theoretical approximation is:
\begin{equation}
\scalemath{0.83}{
    SNR = \frac{P(-1|\alpha=\pi) - P(-1|\alpha=0)}{mean_\alpha \left\{ \sqrt{P(-1|\alpha = x)*[1-P(-1|\alpha = x)]} \right\}} * \sqrt{n}
}
\label{eq:theo_snr}
\end{equation}

\begin{figure}
     \centering
     \begin{subfigure}[b]{0.25\textwidth}
         \centering
         \includegraphics[width=\textwidth]{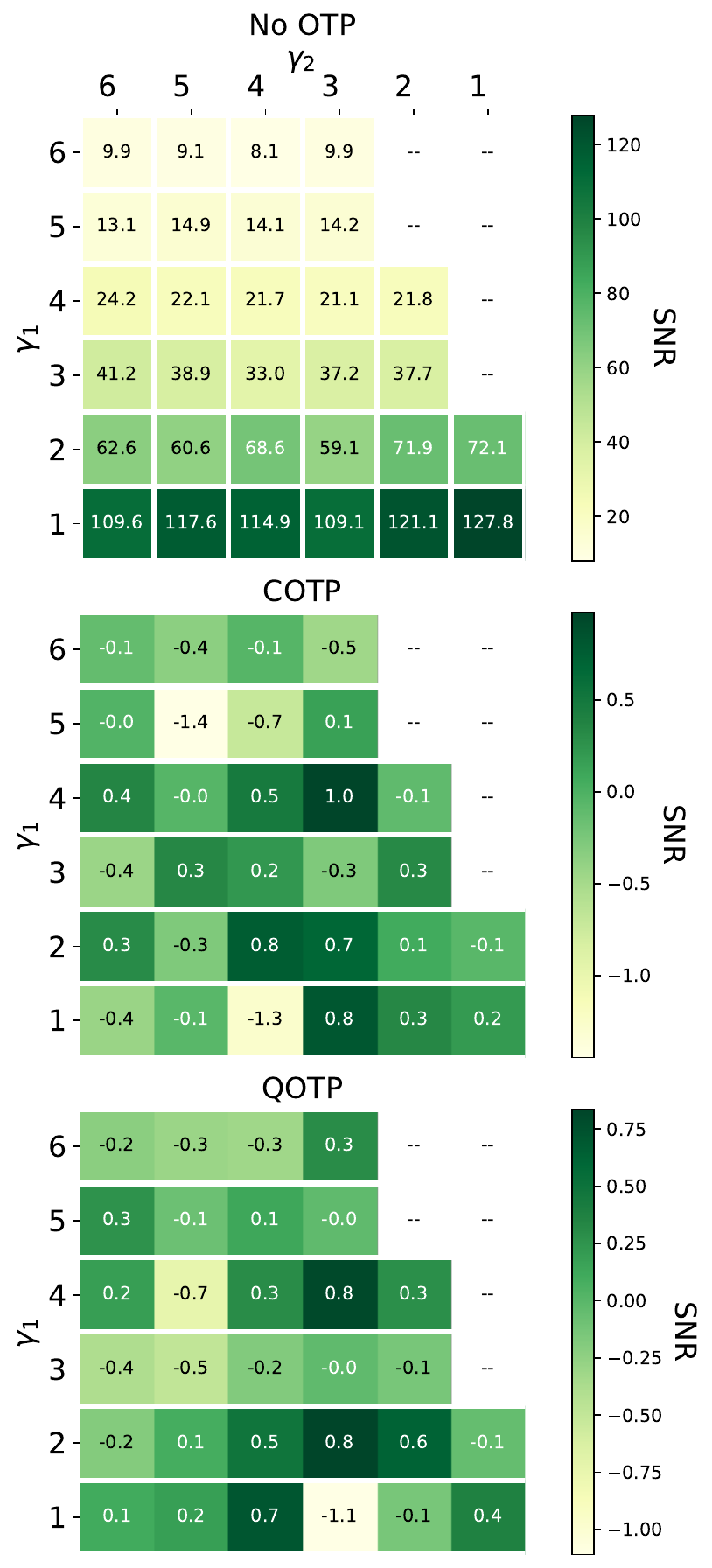}
         \caption{\small Along the $Z$ axis.}
         \label{fig:snr_delay_z}
     \end{subfigure}~
     \begin{subfigure}[b]{0.25\textwidth}
         \centering
         \includegraphics[width=\textwidth]{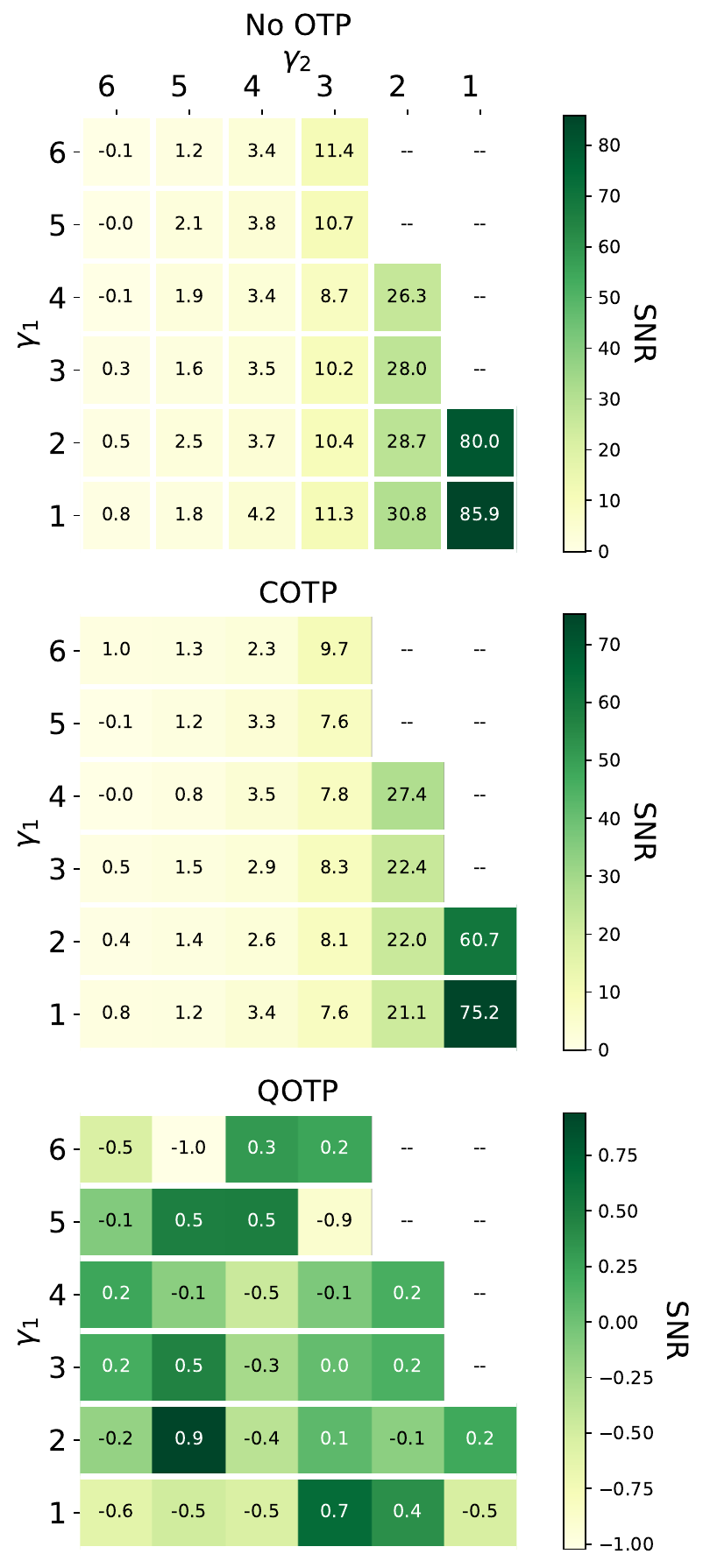}
         \caption{\small Along the $X$ axis.}
         \label{fig:snr_delay_x}
     \end{subfigure}
        \caption{\small SNR on the simulator with different $\gamma_1$ and $\gamma_2$, the ratio of 250 ns to the decoherence time $T_1$ and $T_2$ defined in Section~\ref{sec:thermal}. Note $\gamma_1 \leq 2\gamma_2$ due to $T_2 \leq 2T_1$. (a) The measurement axis is along the $Z$ axis; (b) the measurement axis is along the $X$ axis.}
        \label{fig:snr_delay}
\end{figure}

For the delay, the main factors are $T_1$ and $T_2$, or the corresponding $\gamma_1$ and $\gamma_2$ defined in Section~\ref{sec:thermal}. The results shown in Figure~\ref{fig:snr_delay} is consistent with Section~\ref{sec:thermal}. When measuring along the $Z$ axis, only $\gamma_1$ is important, and when measuring along the $X$ axis, only $\gamma_2$ influences the results. Without OTP, the state leakage is apparent along both axes. COTP can only mitigate state leakage along the $Z$ axis and has a similar pattern as no OTP when measuring along the $X$ axis, while QOTP can mitigate state leakage along all axes.

\begin{figure}
     \centering
     \begin{subfigure}[b]{0.25\textwidth}
         \centering
         \includegraphics[width=\textwidth]{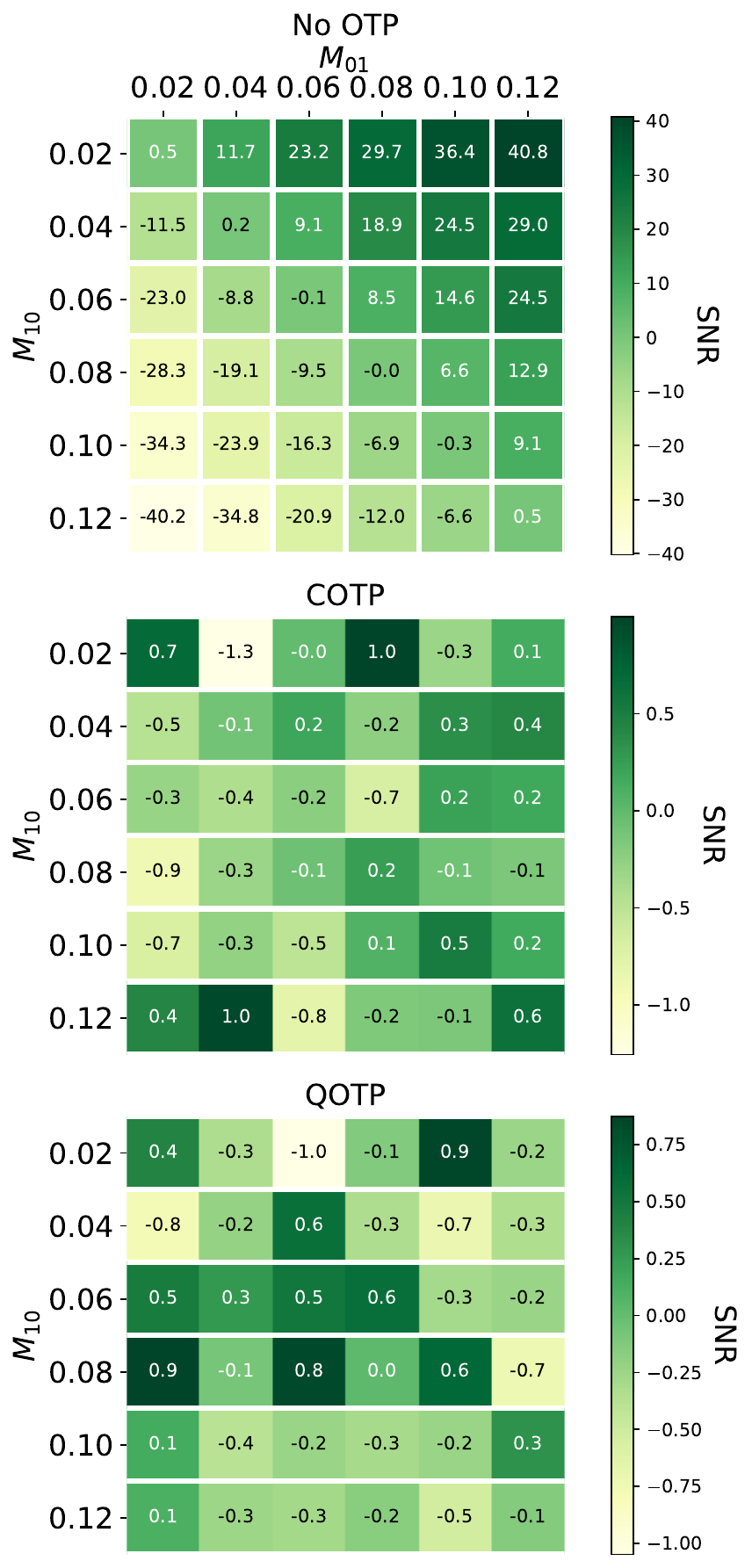}
         \caption{\small Along the $Z$ axis.}
         \label{fig:snr_reset_z}
     \end{subfigure}~
     \begin{subfigure}[b]{0.25\textwidth}
         \centering
         \includegraphics[width=\textwidth]{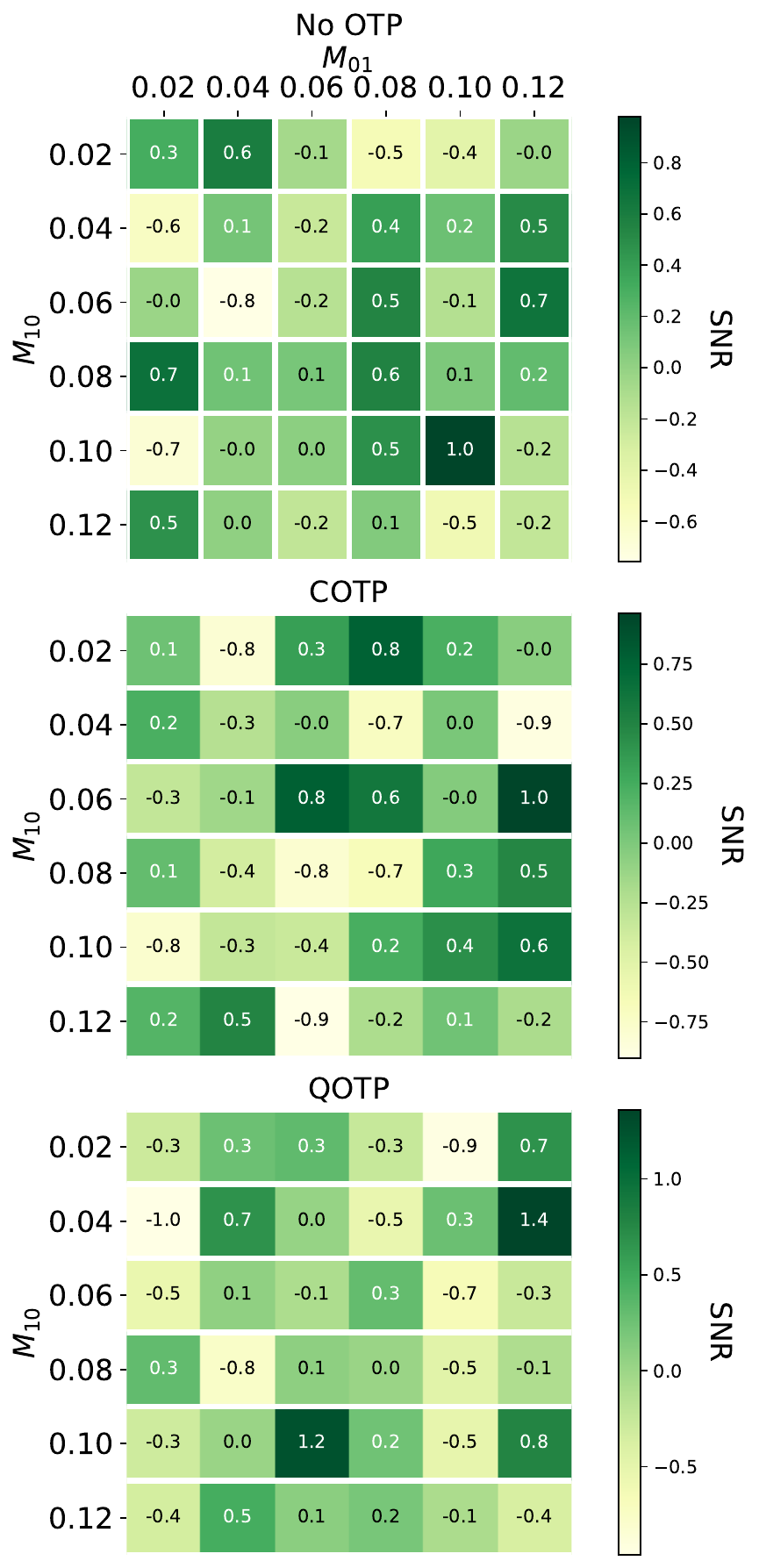}
         \caption{\small Along the $X$ axis.}
         \label{fig:snr_reset_x}
     \end{subfigure}
        \caption{\small SNR on the simulator with different $M_{10}$ and $M_{01}$, the measurement error defined in Section~\ref{sec:reset}. (a) The measurement axis is along the $Z$ axis; (b) the measurement axis is along the $X$ axis.}
        \label{fig:snr_reset}
\end{figure}

For the reset instruction, due to the existence of the measurement that will project all states to its eigenstates, COTP and QOTP will have the same effect if the axis of the measurement and the gate of the OTP are correctly selected, as we listed in the requirements for COTP in Section~\ref{sec:requirements}. The results are shown in Figure~\ref{fig:snr_reset}, and only the measurement error $M_{10}$ and $M_{01}$ are considered in this figure while excluding the error of the conditional Pauli-$X$ for simplicity. The state leakage is not mitigated only when measuring along the $Z$ axis and without OTP, while in other cases the dependence is removed. In addition, from the results of measuring along the $Z$ axis and without OTP, it is proved that SNR is dependent on $M_{01} - M_{10}$: if the measurement error is not much biased among $\ket 0$ and $\ket 1$, then the state leakage is small, as demonstrated in Section~\ref{sec:reset}. Also, the reverse direction of the state leakage pattern is observed with SNR $< 0$ when $M_{10} > M_{01}$.

\begin{figure}
     \centering
     \begin{subfigure}[b]{0.25\textwidth}
         \centering
         \includegraphics[width=\textwidth]{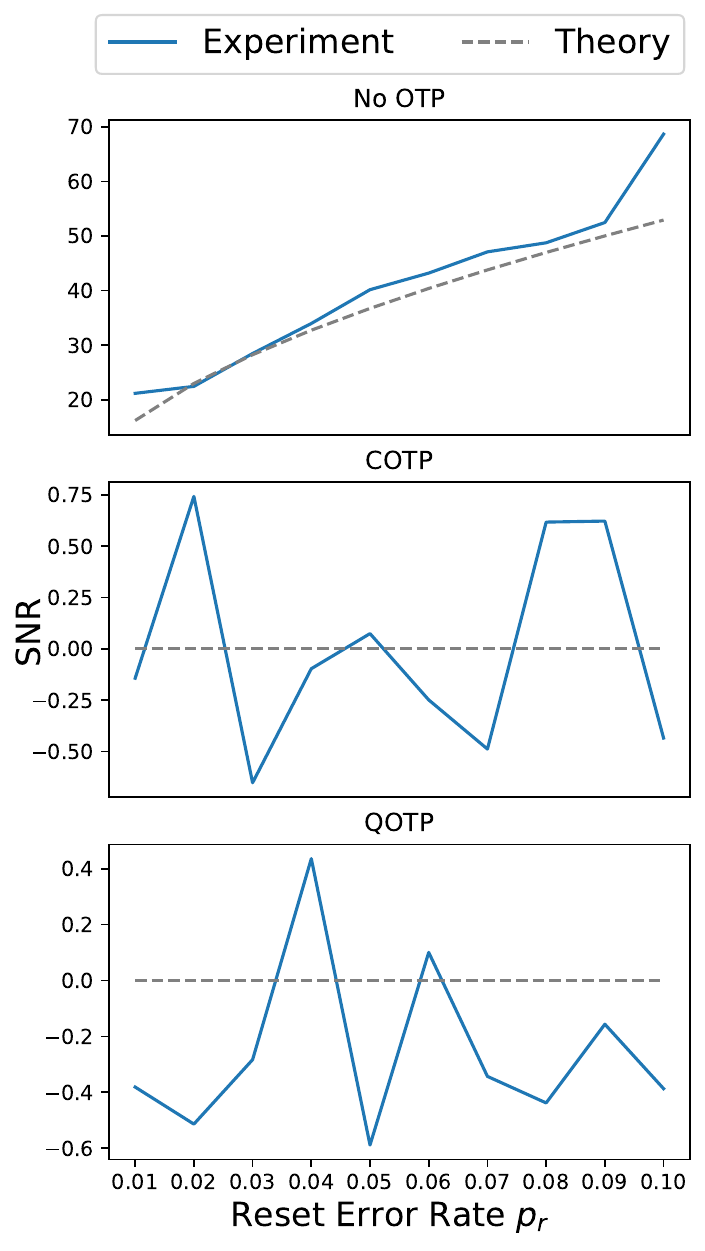}
         \caption{\small Along the $Z$ axis.}
         \label{fig:fig:snr_imag_reset_z}
     \end{subfigure}~
     \begin{subfigure}[b]{0.25\textwidth}
         \centering
         \includegraphics[width=\textwidth]{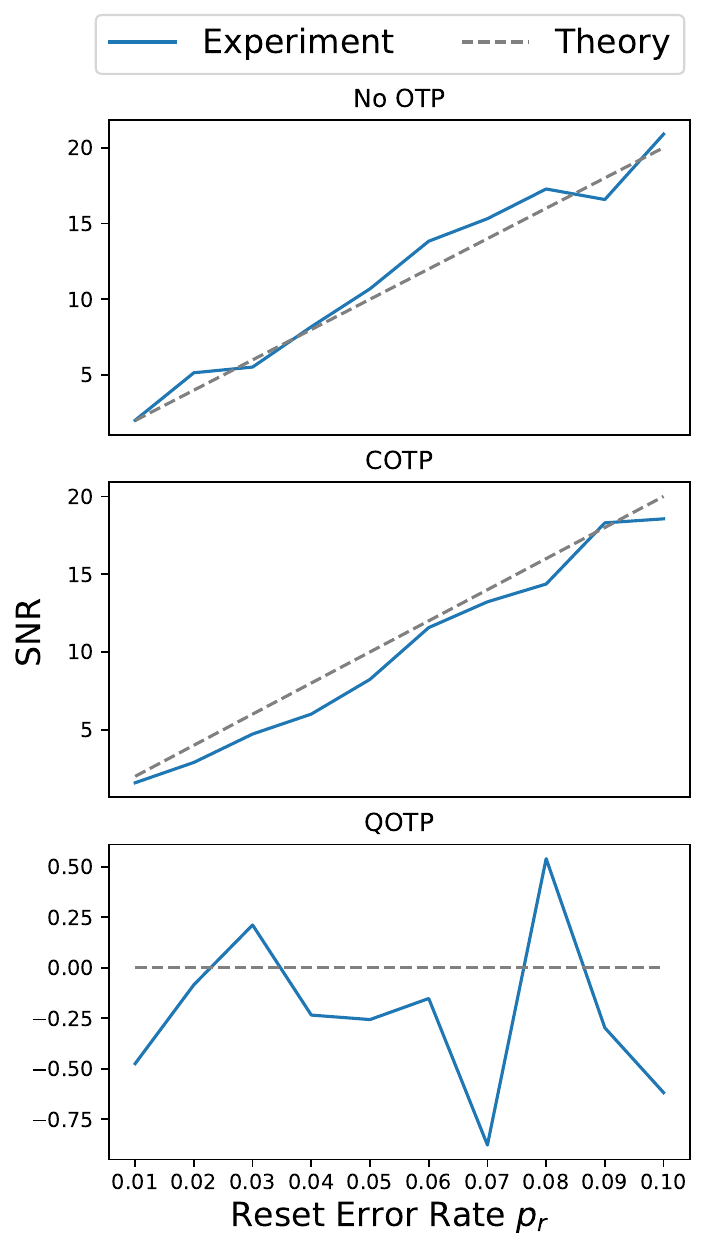}
         \caption{\small Along the $X$ axis.}
         \label{fig:fig:snr_imag_reset_x}
     \end{subfigure}
        \caption{\small SNR on the simulator with different $p_r$, the error of the measurement-less reset instruction defined in Section~\ref{sec:imag_reset}). (a) The measurement axis is along the $Z$ axis; (b) the measurement axis is along the $X$ axis.}
        \label{fig:snr_imag_reset}
\end{figure}

Last, for the measurement-less reset instruction, which is a simplified and isotropic version of the delay, only the reset instruction error $p_r$ plays a role. As shown in Figure~\ref{fig:snr_imag_reset}, similar to the delay, COTP can only mitigate state leakage when along the $Z$ axis and is nearly the same as the case without OTP. On the contrary, QOTP can mitigate state leakage along all axes.

Besides the noise and errors in quantum computers themselves, one of the most important factors is the number of shots of the attacker measurement. According to Equation~\ref{eq:theo_snr}, the attacker can easily increase SNR by increasing the number of shots. In theory, this means any non-zero dependence of the probability on the input parameter can be measured by attackers. Unless the design fully mitigates the state leakage, such as the QOTP and COTP with the reset instruction, the state leakage may be abused. Nonetheless, this assumption is based on the hardware being in the same condition. For example, the noise and error models should be the same across all shots. Such a requirement is unrealistic in NISQ quantum computers, so the state leakage is not extraordinary or feasible to be detected when it is small, such as shown in Figure~\ref{fig:qc_delay_x}.

\section{Conclusion}
\label{sec:conclusion}
This study examines the state leakage problem in quantum computing and suggests using the one-time pad before the reset operations to mitigate state leakage. Though the classical one-time pad cannot mitigate state leakage in most cases, this study examines the prerequisites for it to work and shows that its synergy with reset instruction can be a more economical substitution for the quantum one-time pad. By comparing degrees of leakage under various levels of error rates, this paper evaluates the role of errors in state leakage. New insights on the creation of safe quantum computing systems and reset procedures are provided by our findings.

\clearpage

\bibliographystyle{IEEEtran}
\bibliography{bibliography.bib}

\end{document}